\documentclass[final,3p,twocolumn]{elsarticle}

\makeatletter
\def\ps@pprintTitle{%
  \let\@oddhead\@empty
  \let\@evenhead\@empty
  \let\@oddfoot\@empty
  \let\@evenfoot\@oddfoot
}
\makeatother

\usepackage{lineno}
\modulolinenumbers[5]
\usepackage{hyperref}
\usepackage{svg}
\hypersetup{colorlinks=true,pdfborder=0 0 0}
\usepackage{subcaption}
\usepackage{amsmath,bm,siunitx,amssymb,caption,booktabs}

\usepackage[normalem]{ulem} 

\usepackage{xspace}

\usepackage{textcomp,gensymb}
\usepackage{cleveref}
\usepackage{xcolor}

\newcommand{\tool}{GibbScope\xspace}
\newcommand{\rerun}[1]{#1}

\journal{arxiv}

\begin{document}

\begin{frontmatter}

\title{GibbScope: interactive calculation and visualization of \\ 
phase diagrams in the browser}

\address[az-cs]{Department of Computer Science, University of Arizona, Tucson, AZ 85721, USA}
\address[az-mse]{Department of Materials Science and Engineering, University of Arizona, Tucson, AZ 85721, USA}
\address[az-am]{Graduate Interdisciplinary Program in Applied Mathematics, University of Arizona, Tucson, AZ 85721, USA}

\author[az-cs]{Ramya Ramachandran}
\author[az-mse]{Zhuocheng Huang}
\author[az-mse]{Jiayang Wang}
\author[az-mse,az-am]{Marat I. Latypov\corref{cor1}}
\cortext[cor1]{corresponding author}
\ead{latmarat@arizona.edu}

\begin{abstract}

Phase diagrams are a constant work in progress as Gibbs-energy models are fitted, revised, and disputed in the light of an endless stream of experimental and computational data. Yet, phase diagrams are still largely communicated as static figures, in much the same way as a century ago. We present \tool, a web-based framework for interactive calculation and visualization of binary phase diagrams. Written in Rust and compiled to WebAssembly, the new CALPHAD engine recomputes a complete phase diagram in the browser in a fraction of a second with zero installation on the user or server side. The high speed allows new ways of interacting with phase diagrams and underlying assessments. Validation against well-established CALPHAD software shows agreement to within \qty{0.01}{\kelvin} for invariant temperatures and within \num{1e-6} for tie-line compositions for most databases. GibbScope offers interactive CALPHAD parameter adjustments, overlays of different assessments and point-by-point misfit to experimental data, which makes the diagram a live, working object. In addition, \tool provides convenient access to \rerun{\num{2745}} thermodynamic databases from the SGTE collection (Hallstedt, 2025) and their live adjustment. For \rerun{\num{1254}} of these databases, all phases are modeled in full with the current implementation. \\ 
GibbScope is freely available at \url{https://gibbscope.github.io}.

\end{abstract}


\end{frontmatter}

\section{Introduction}
\label{sec:intro}

Phase diagrams are a cornerstone of materials science and a constant work in progress. Gibbs-energy models are fitted, revised, and disputed in the light of an endless stream of experimental and computational data. Yet, binary phase diagrams are still largely communicated as static figures, pretty much the same way as a century ago. A lot has changed over the same period, ranging from the way we construct phase diagrams to how we interact with data.

First, phase diagrams are now routinely calculated instead of being drawn directly from experimental measurements. The CALPHAD (CALculation of PHAse Diagrams) method \cite{kaufman1970,lukas2007} allows for computing phase diagrams from thermodynamic models of individual phases. In the CALPHAD approach, the Gibbs energy of each phase is described as a function of temperature and composition with model parameters fitted to experimental phase equilibria and thermochemical data, often complemented by first-principles calculations \cite{liu2009}. Since its introduction \cite{kaufman1970}, the CALPHAD method has been applied to thousands of systems and has become a key tool in computational materials design \cite{olson1997,kattner2016}. However, phase diagrams computed with CALPHAD are still exported, published, and used as static images. 

Static phase diagrams give the impression of an ultimate truth set in stone, especially to students reading phase diagrams from a textbook. Yet, a phase diagram is a display of an individual prediction with a single set of model parameters calibrated to a particular dataset. As such, calculated phase diagrams are living and evolving: assessments are revised as new experimental data, first-principles calculations, and thermodynamic models become available. In addition, the same data may have different interpretations, which results in different phase diagrams of the same system. For example, the recently published SGTE collection of \num{2773} binary datasets \cite{hallstedt2025} for \num{1241} binary systems contains \rerun{\num{598}} systems with two or more published assessments, and as many as 16 assessments for systems like Fe--Nb and Fe--Ti. Different assessments of the same system can differ in invariant points, homogeneity ranges of intermediate phases, and even in the phase diagram topology (\Cref{sec:living}). In addition to the differences between assessments, the parameters are subject to uncertainty. Recent work  has shown that the uncertainty in model parameters propagates into appreciable uncertainty of phase boundaries and invariant reactions  \cite{stan2003,otis2017jom,paulson2019,honarmandi2019}. A static figure of a single assessment shows neither of these aspects, and examining them requires recomputing and comparing phase diagrams for different assessments or parameter sets.

This limitation may come as a surprise as there is no shortage of modern software in the field. Commercial software packages such as Thermo-Calc \cite{andersson2002}, FactSage \cite{bale2016}, Pandat \cite{cao2009}, and MTDATA \cite{davies2002} provide robust calculations of phase equilibria in multicomponent systems along with curated thermodynamic databases. Relevant open-source projects include PyCalphad \cite{otis2024pycalphad}, OpenCalphad \cite{sundman2015}, and ESPEI \cite{bocklund2019}. These tools require installation and either licenses for commercial packages or programming skills for open-source libraries. Several web-based services have recently made CALPHAD calculations accessible through the browser (e.g., \cite{yan2026phase,nanohubtc}). In these services, however, calculations are performed on remote servers, which requires uploading user data and a round trip to the server for each change in a thermodynamic database, calculation parameter, or visualization setting. Users who only need to read a phase diagram typically rely on static compilations such as the ASM Alloy Phase Diagram Database \cite{okamoto1990binary} or, most recently, on a collection of precomputed phase diagrams served together with literature data \cite{arroyave2026orchid}.

\begin{figure*}[t]
  \centering
  \includegraphics[width=0.9\textwidth]{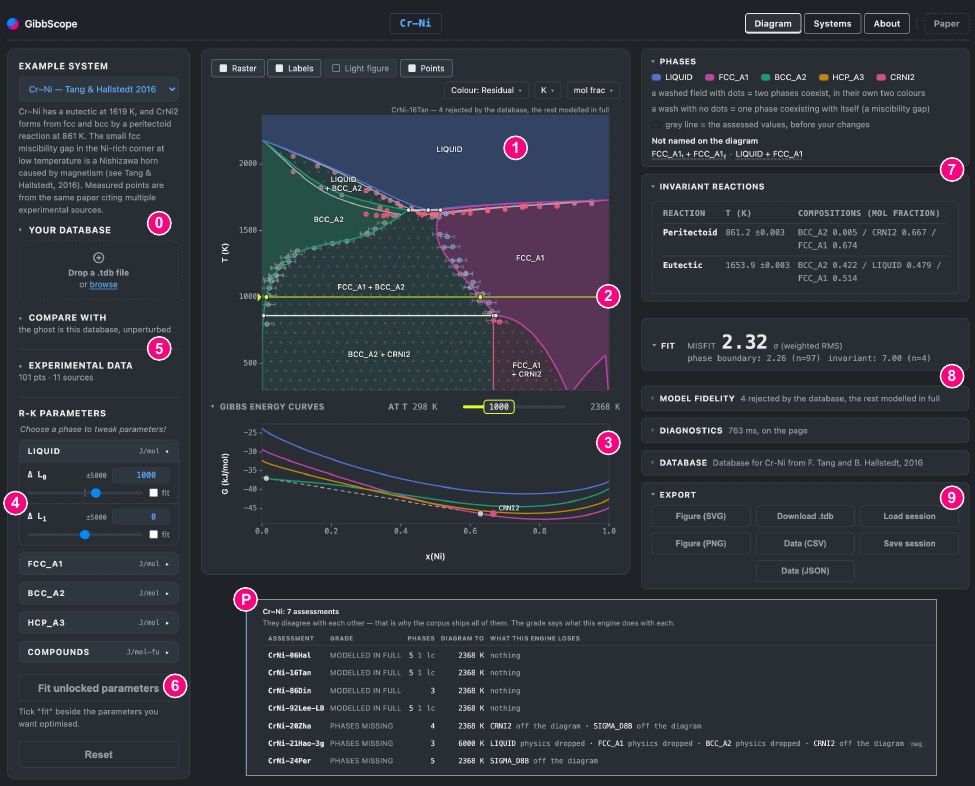}
    \caption[\tool at a glance]{\tool at a glance. (0) Loading of the user's own \texttt{.tdb} file. (1) Phase diagram. (2) Isotherm at \qty{1000}{\kelvin}. (3) Gibbs energy curves at the isotherm; the dashed common tangent of b.c.c.\ and f.c.c.\ gives the tie line in \textcircled{2}. (4) Parameter sliders: $L_0$ of the liquid is increased by \qty{1}{\kilo\joule\per\mole}, which raises the eutectic from \rerun{\num{1619}} to \rerun{\qty{1654}{\kelvin}} (gray lines: assessed diagram). (5) Comparison with another assessment and with experimental data; the measured points used in the assessment \cite{tang2016using} are colored by their signed residual (red: model above the point; blue: below). (6) Fitting of selected parameters to the data. (7) Legend of the phases and invariant reactions. (8) Misfit and model fidelity summary. (9) Export of figures, data, the modified database, and the session. Inset (P) Assessments of Cr--Ni in the SGTE collection with their grades, listed in the Systems tab (\Cref{fig:picker}). The interface shows the Cr--Ni assessment \cite{tang2016using}, which is a default preloaded example in the app. The interface view can be reopened at \url{https://gibbscope.github.io/?example=crni-16tan&col=residual&p=LIQUID:0:a=-12300}.}
  \label{fig:concept}
\end{figure*}

Meanwhile, the way we interact with data has dramatically changed over the years. Interactive web applications have become a common way of exploring information \cite{bostock2011d3}, including materials data  (e.g., the Materials Project \cite{horton2025accelerated}, Materials Cloud \cite{talirz2020materials}, or BisQue \cite{latypov2019bisque,polonsky2019three}). New technologies such as WebAssembly allow web browsers to execute compiled code at near-native speed \cite{haas2017}, which opens opportunities for interactive experiences even in applications with demanding calculations without a server \cite{perkel2024no}. Visualization and analysis of phase equilibria could clearly benefit  \cite{liu2014effects} from a tool that combines the rigor of CALPHAD calculations from assessed databases, the accessibility of a web page, and the near-native speed needed for interactive exploration. To our knowledge, such capabilities have not been reported for CALPHAD to date.

Here, we present \tool, a new framework for interactive calculation and visualization of binary phase diagrams in the browser. \tool is based on a CALPHAD engine written in Rust and compiled to WebAssembly, which runs entirely in the user's web browser with zero installation on the local computer or server. The engine recomputes a complete binary phase diagram in a fraction of a second, which allows the user to change model parameters and follow the response of the phase diagram live, overlay competing assessments of the same system, and evaluate the agreement between the computed diagram and experimental data (\Cref{fig:concept}). Below, we first describe the user interface of \tool and interactions with phase diagrams (\Cref{sec:interface}) as well as the implementation choices that make this interactivity possible (\Cref{sec:framework}). We then validate the engine against PyCalphad and Thermo-Calc (\Cref{sec:validation}), and apply it to the entire SGTE collection of binary datasets (\Cref{sec:corpus}). Finally, we use the speed of the engine to quantify how much assessed phase diagrams change with their model parameters and to rapidly inspect differences between assessments of the same system (\Cref{sec:living}).

\begin{figure*}[t]
  \centering
  \includegraphics[width=\textwidth]{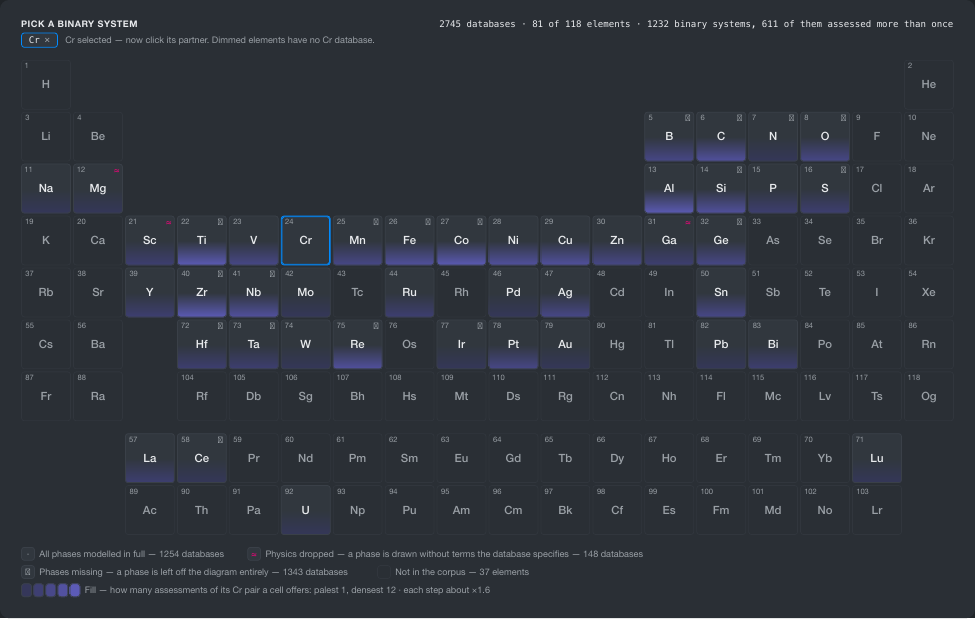}
  \caption[Picker]{Interactive picker of binary systems in \tool, shown in its periodic-table view after an element (Cr) is selected. Elements that form a binary system with Cr in the SGTE collection are highlighted with the others dimmed. The fill of each cell gives the number of Cr--X databases (from 1 to 12), and the mark in its corner gives the grade of the best of them (no mark: all phases modeled in full; $\approx$: physics dropped; $\boxtimes$: phases missing). Selecting the partner element lists the assessments of the system with their grades (inset \textcircled{P} in \Cref{fig:concept}). This view is accessible online at \url{https://gibbscope.github.io/?tab=picker}.}
  \label{fig:picker}
\end{figure*}

\section{Results}
\label{sec:results}

\subsection{User interface and interactive features of \tool}
\label{sec:interface}

\Cref{fig:concept} shows the user interface of \tool developed in this work. The interface consists of a single  web page with three major components: user input (left column), a canvas for the phase diagram and Gibbs energy curves (center column), and a summary of the results with export options (right column). The user first loads a thermodynamic database to work with by either selecting one of the preloaded examples, dropping their own \texttt{.tdb} file into the page (\textcircled{0}), or loading a database from the SGTE collection \cite{hallstedt2025} bundled with the application. The \rerun{\num{2745}} bundled databases of the collection are accessible through the system picker (``Systems''), which opens an interactive periodic table (\Cref{fig:picker}). 

Once a database is loaded, the calculated phase diagram is displayed in the central canvas (\textcircled{1}), which supports zooming into regions of interest. Hovering over a point of the diagram shows a tooltip with the stable phase or, in a two-phase field, the two coexisting phases, the tie line through the point, and the phase fractions given by the lever rule. Temperatures above the ranges declared in the database for the stable phases are hatched to mark extrapolation (\Cref{sec:methods-sweep}). A movable isotherm (\textcircled{2}) sets the temperature of the panel below the diagram, which shows the Gibbs energy curves of all phases and their lower convex hull at this temperature (\textcircled{3}). The two plots share the composition axis, so that each common tangent of the Gibbs energy curves lines up with a tie line of the phase diagram. Changing the temperature of the isotherm interactively updates the Gibbs energy curves, which connects each isotherm of the phase diagram with the underlying Gibbs energies.

One of the central features of \tool is the live adjustment of model parameters (\textcircled{4}). The constant term of each Redlich--Kister parameter and the Gibbs energy of each line compound can be changed with a slider, which controls the change from the assessed value. The phase diagram is recomputed as the slider moves, while the assessed diagram remains in the background as gray lines. The user can therefore see at once which phase boundaries and invariant reactions respond to a parameter and by how much. In \Cref{fig:concept}, an increase of the zeroth-order parameter of the liquid by \qty{1}{\kilo\joule\per\mole} raises the eutectic temperature of Cr--Ni by \rerun{\qty{35}{\kelvin}} (see further analysis with this capability in \Cref{sec:living}).

The displayed phase diagram can be overlaid with experimental data or another phase diagram (\textcircled{5}). The ``Compare with'' feature plots a phase diagram of another assessment of the same system in the background as neutral gray lines in place of the assessed diagram, so that differences in phase boundaries, invariant temperatures, and topology between the two assessments can be easily seen (\Cref{sec:living}). Experimental phase equilibria can be loaded from CSV, TSV, or JSON files or from the data formats of PARROT and ESPEI. The points are drawn on the phase diagram and colored by their signed residual, i.e., their distance to the computed diagram scaled by their uncertainties. The misfit to all points is displayed (\textcircled{8}). Parameters selected by the user can also be fitted to the data (\textcircled{6}), which provides a starting point for a full assessment (\Cref{sec:methods-misfit}).

The right column summarizes the displayed diagram. It contains a legend of the phases, a table of the invariant reactions found by the engine with their types, temperatures, and phase compositions (\textcircled{7}; \Cref{sec:methods-sweep}), the misfit to experimental data and the fidelity notes (\textcircled{8}), diagnostics with the calculation time, the comments of the \texttt{.tdb} file, and the export options (\textcircled{9}). The legend also lists the phase fields that are too small to be labeled in the diagram, and hovering over an entry of this list highlights the corresponding field.

Every view in the application can be saved, shared, and exported (\textcircled{9}). The state of the application (selected database, compared assessment, axis ranges, temperature, display settings, and any modified parameters) is encoded in the URL, so that a link reopens the same view of a database from the bundled collection, including the modified phase diagram, in any browser (see the link in the caption of \Cref{fig:concept}). Experimental data are not included in the URL but can be saved together with the session in a file. The phase diagram can be exported as a figure (PNG or SVG) or data (CSV or JSON), and the modified database as a \texttt{.tdb} file, in which the changed parameters are written into the corresponding \texttt{PARAMETER} statements and the rest of the file remains unchanged. The exported database can therefore be used in established tools for further calculations or a full assessment.

The system picker for the bundled SGTE collection is accessible via ``Systems'' tab and is also interactive. Once the first element is selected, the partner elements that form a binary system with it are highlighted according to the number of available databases (\Cref{fig:picker}). Selecting the second element lists all assessments of the pair with their grades and the phases that the engine cannot model (\textcircled{P} in \Cref{fig:concept}). The phase diagram of the selected assessment is computed and displayed within a fraction of a second. Since the engine does not implement all thermodynamic models used in the collection, a fidelity summary reports whether the phases are modeled in full, computed with some terms omitted, or missing entirely (\textcircled{8}; \Cref{sec:framework}). The same summary is printed above the diagram and written into exported figures.

The interactivity and responsiveness of the application are enabled by a special implementation of the CALPHAD engine focused on computational efficiency as described next.

\subsection{Interactive CALPHAD in the browser with Rust and WebAssembly}
\label{sec:framework}

The \tool framework consists of two main components: (i) a CALPHAD engine that performs all thermodynamic calculations and (ii) a web application that handles the user interaction and visualization. The engine is written in Rust \cite{matsakis2014rust} and compiled to WebAssembly -- a binary instruction format that modern web browsers execute at near-native speed \cite{haas2017}. Compiling the engine to WebAssembly allows all calculations to run in the user's browser instead of a remote server. As a result, \tool requires no installation, and no data leave the user's computer, including databases, modified parameters, or experimental data dropped into the interface. The browser only needs to download the engine once (\rerun{\qty{0.8}{\mega\byte}}, or \rerun{\qty{0.3}{\mega\byte}} compressed) along with the database selected by the user (a text file of several kB). Every change made by the user is then computed on the user's own processor, and the calculations continue to work even without a network connection once the page and the database are loaded.  Since the server only delivers static files, \tool does not require computing resources on the server side, regardless of the number of users (which allows hosting on GitHub Pages). Such client-side computing is only beginning to emerge in computational materials science, and, to our knowledge, \tool is the first CALPHAD engine that runs entirely in the browser.

\begin{table*}[t]
  \centering
  \caption{Summary of the comparisons of the \tool engine with PyCalphadand Thermo-Calc. $\Delta T$ and $\Delta x$ are absolute differences in temperature and mole fraction, respectively.} 
  \label{tab:validation}
  \small
  \begin{tabular}{llrrll}
    \toprule
    Reference & Quantity & Systems & Comparisons & $\Delta T$ (K) & $\Delta x$ \\
    \midrule
    PyCalphad & invariants, curated & 10 & 17 & max \rerun{\num{0.001}} & $<$\rerun{\num{5e-5}} \\
    PyCalphad & invariants, random sample & 80 & \rerun{\num{383}} of 384 & median \rerun{\num{0.0009}} & median \rerun{\num{8.6e-7}} \\
    PyCalphad & Gibbs energy, magnetic term & 11 cases & 561 & \multicolumn{2}{l}{max \rerun{\num{1.5e-6}} relative} \\
    Thermo-Calc & invariants & 42 & 217 & max \rerun{\num{0.0052}} & max \rerun{\num{3.1e-5}} \\
    Thermo-Calc & equilibria at fixed $T$, $x$ & 48 & \num{1897} & -- & max \rerun{\num{7.8e-7}} \\
    Thermo-Calc & mapped boundaries & 48 & \rerun{\num{12051}} of \num{12152} & -- & median \rerun{\num{3.8e-8}} \\
    \bottomrule
  \end{tabular}
\end{table*}

The unique features of \tool are enabled by the implementation of CALPHAD calculations in Rust, which was chosen for two reasons. First, compiled Rust code runs at a speed comparable to that of C and C++, while its memory safety is checked by the compiler and memory is managed without a garbage collector. As a result, calculations are not interrupted by pauses for memory management when the user interacts with the code (e.g., changes a parameter by dragging a slider). Second, the same source code is compiled both to native machine code and to WebAssembly. These compilation options allow running the same code natively for testing and as WebAssembly in the browser. In this work, this feature is leveraged for validation against established CALPHAD software (\Cref{sec:validation}). 

The speed achieved by the Rust implementation and compilation into WebAssembly enables the key feature of \tool: the interactivity of a phase diagram. A thermodynamic database (\texttt{.tdb}) is parsed once into polynomial coefficients of the Gibbs energy of each phase. A subsequent change of a model parameter in the user interface replaces a single coefficient without parsing the database again (\Cref{sec:methods-parsing}). A complete phase diagram, with all phase boundaries and invariant reactions over the default ranges of composition and temperature, is then computed in a fraction of a second. Tests carried out on thousands of binary phase diagrams from the SGTE collection show a median calculation time of \rerun{\qty{0.35}{\second}} and at most \rerun{\qty{1.2}{\second}} on a consumer-grade laptop (see \Cref{sec:methods-timing}). Such short calculation times allow the user to adjust the parameters of the Gibbs free energy functions and watch nearly instantaneous updates in the phase diagram according to the parameter change. During such interactive adjustments, the figure shows the last completed calculation until the next one is finished. Since a typical calculation takes a fraction of a second, the delay is barely noticeable and does not degrade the responsiveness and the user experience.  

Under the hood, the engine follows a typical CALPHAD workflow for binary systems: it parses the database, evaluates the Gibbs energies of the phases, finds the stable phases at each temperature from the lower convex hull of the Gibbs energy curves \cite{gibbs1876}, and scans the temperature to trace the phase boundaries, to locate and classify the invariant reactions (\Cref{sec:methods}). Solution phases are treated as substitutional solutions of the two elements, and their Gibbs energies include SGTE unary reference energies \cite{dinsdale1991}, ideal mixing, Redlich--Kister excess energy \cite{redlich1948}, and the Inden--Hillert--Jarl magnetic contribution \cite{inden1981,hillert1978}. Line compounds are described with a single Gibbs energy per formula unit. Phases that require other models (e.g., compound energy formalism with multiple sublattices, ionic liquid, or gas) are not supported in the current implementation. To reflect the limitations, the engine grades each phase declared in the database  as \textit{modeled in full} (every term given in the database is evaluated), \textit{physics dropped} (a term given in the database is omitted, e.g., an associate species or a third-generation parameter), \textit{phase missing} (the phase is left out of the diagram), or \textit{rejected by the database} (the phase is excluded by the database itself). The grades are reported in the caption of the phase diagram, both in the application and in exported figures, so that the user is aware of the fidelity of the displayed diagram.

The web application communicates with the engine through a JavaScript interface generated with \texttt{wasm-bindgen} \cite{wasmbindgen}, so that the engine appears to the application as an ordinary JavaScript module. To keep the user interface responsive, the phase diagram is computed in a background thread (Web Worker) with its own instance of the engine, while a second Web Worker is used for fitting model parameters to experimental data. 

\subsection{Validation} 
\label{sec:validation}

\begin{figure*}[t]
  \centering
  \includegraphics[width=\textwidth]{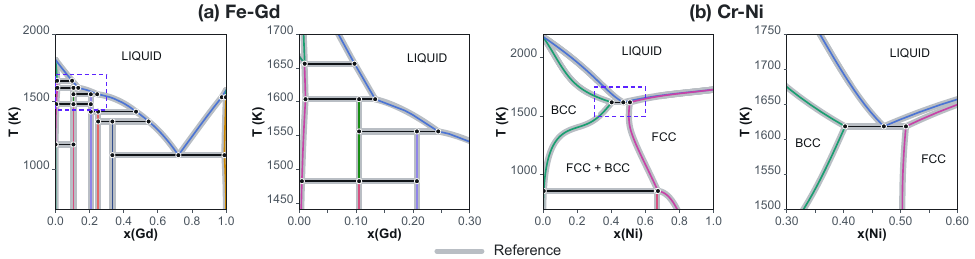}
  \caption[Validation]{Validation of the \tool engine against Thermo-Calc
  for (a) Fe--Gd \cite{chen2017} and (b) Cr--Ni \cite{tang2016using}. Phase boundaries computed with \tool (lines colored by phase as in the application, e.g., blue: liquid, green: b.c.c., magenta: f.c.c.) and the compositions of the phases in invariant reactions (black dots) are drawn over the phase diagram mapped by Thermo-Calc from the same database (thick gray lines, ``Reference''). The right panel of each pair enlarges the region in the dashed box: the cascade of invariant reactions at $x(\mathrm{Gd}) < 0.3$ in Fe--Gd and the eutectic in Cr--Ni. The \tool layer of each panel can be reopened in the online application for both Fe--Gd (\href{https://gibbscope.github.io/?db=FeGd-17Che&x=0,1&t=700,2000&g=0&fig=light&mode=vector}{full}, \href{https://gibbscope.github.io/?db=FeGd-17Che&x=0,0.3&t=1440,1700&g=0&fig=light&mode=vector}{enlarged}) and Cr--Ni (\href{https://gibbscope.github.io/?db=CrNi-16Tan&x=0,1&t=700,2200&g=0&fig=light&mode=vector}{full}, \href{https://gibbscope.github.io/?db=CrNi-16Tan&x=0.3,0.6&t=1500,1750&g=0&fig=light&mode=vector}{enlarged}) via hyperlinks in the parentheses.}
  \label{fig:validation}
\end{figure*}

The \tool engine was validated against two well-established (\emph{reference}) implementations of CALPHAD: PyCalphad \cite{otis2024pycalphad} and Thermo-Calc \cite{andersson2002}. Validation was performed by feeding the same databases to the three programs. For one-to-one comparison, phases that are currently unsupported by \tool (\Cref{sec:corpus}) were excluded from the reference calculations so that the comparison focuses on numerical accuracy instead of differences in the supported models or phase types. The validation shows excellent agreement both visually (\Cref{fig:validation}) and numerically (\Cref{tab:validation}). 

Visual comparison was performed by overlaying phase diagrams from \tool and Thermo-Calc. \Cref{fig:validation} shows that the phase boundaries from the two programs coincide for two representative systems: Fe--Gd with a cascade of invariant reactions involving line compounds \cite{chen2017} and Cr--Ni with a eutectic between the b.c.c.\ and f.c.c.\ solutions \cite{tang2016using}.

Numerical validation ensured that the agreement extends to a wide range of phase diagrams of various topologies. To this end, we sampled dozens of phase diagrams from the SGTE collection and compared the invariant points, tie lines, and stable phases. For a random sample of 80 databases, \tool reproduced \rerun{\num{383}} of the 384 invariant reactions found with PyCalphad with median differences of about \rerun{\qty{0.001}{\kelvin}} in temperature and \rerun{\num{1e-6}} in the compositions of the participating phases. The one  invariant reaction without a match is a mapping artifact. For a set of 48 systems, comparison with Thermo-Calc showed an agreement on 217 invariant temperatures within \qty{0.01}{\kelvin} as well as the same stable phases identified at a sample of \num{1897} points with tie-line compositions calculated within \num{1e-6}. In addition, the engine matched \rerun{\num{12051}} of the \num{12152} points of the phase boundaries mapped by Thermo-Calc with a median difference of \rerun{\num{3.8e-8}} in composition; the other \rerun{\num{101}} points lie on lines that are metastable in the maps of Thermo-Calc or at the ends of boundary lines. The comparisons and numerical results are summarized in \Cref{tab:validation} and described in \Cref{sec:methods-verification}. 

Overall, the engine reproduces two independent implementations to within millikelvins in temperature and \num{e-6} in composition for most tested phase diagrams, which is well below both the resolution of a phase diagram figure and the typical uncertainty of experimental phase equilibria. 

\subsection{Application to the SGTE collection of binary datasets}
\label{sec:corpus}

\begin{figure*}[t]
  \centering
  \includegraphics[width=\textwidth]{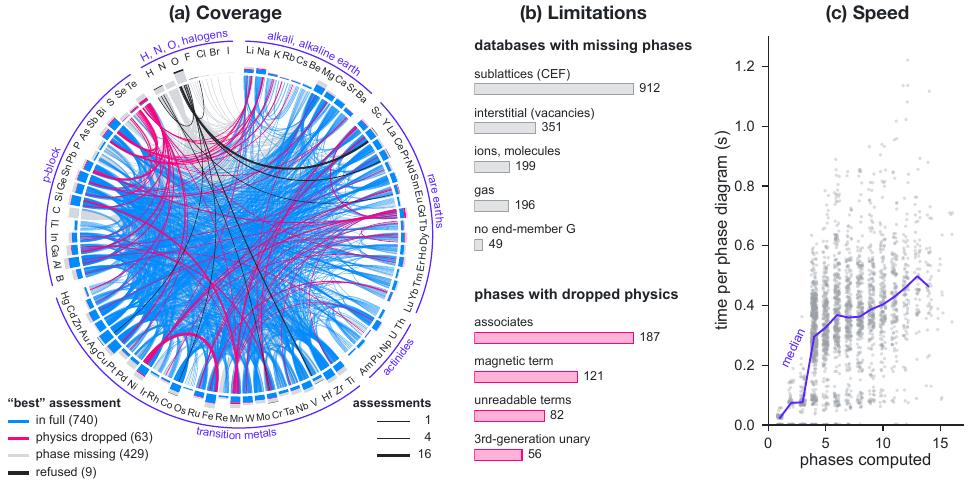}
  \caption[Corpus]{Application of \tool to the SGTE collection of binary datasets \cite{hallstedt2025}: (a) coverage of the binary systems: each chord connects the two elements of a system, its color gives the grade of the best assessment of the system (modeled in full: \rerun{\num{740}} systems; physics dropped: \rerun{\num{63}}; phases missing: \rerun{\num{429}}; refused: \rerun{\num{9}}), and its width gives the number of assessments (1 to 16); the elements are grouped by family, and the outer ring gives the number of systems of each element by grade; (b) reasons for missing phases (counted in databases, under the first applicable reason, \Cref{sec:methods-parsing}) and dropped physics (counted in phases); (c) time to compute a complete phase diagram in Node.js vs.\ the number of phases computed, for each loaded database (dots), with the median (line).}
  \label{fig:corpus}
\end{figure*}

In addition to fast visualization of individual thermodynamic databases and their interactive adjustments, \tool provides convenient access to the SGTE collection of binary assessments compiled by Hallstedt \cite{hallstedt2025}. \tool is bundled with \rerun{\num{2745}} databases out of \num{2773} total. The other \rerun{\num{28}} databases were excluded because \tool does not currently describe any of their phases. The excluded databases describe oxygen, hydrogen, and nitrogen systems  with ionic or gas models (e.g., Y--O, Zr--O, Cr--H, Al--N).
The bundled databases are accessed through the periodic-table picker described in \Cref{sec:interface} (\Cref{fig:picker}).

The \rerun{\num{2745}} bundled databases collectively declare \rerun{\num{22266}} phases. The current implementation of \tool models \rerun{\num{17927}} phases in full and \rerun{\num{404}} phases with ``physics dropped''. The remaining phases are not displayed in their diagrams because they are either unsupported (\rerun{\num{3466}}) by the current engine or rejected by the databases themselves (\rerun{\num{469}}). Graded by their least supported phase, \rerun{\num{1254}} databases (\rerun{\num{46}\%}) are modeled in full, \rerun{\num{148}} have physics dropped in at least one phase without any phase missing, and \rerun{\num{1343}} have at least one phase missing. At the level of systems, \rerun{\num{740}} of the \num{1241} systems in the collection have at least one assessment modeled in full, and \rerun{\num{803}} have at least one assessment without missing phases. These limitations are indicated in the interface when the user is picking elements (\Cref{fig:picker}). A visual summary of the coverage of the SGTE collection and limitations is presented in \Cref{fig:corpus}(a,b). Such analysis is instructive for guiding further development of \tool or similar software: the reasons for missing phases and dropped physics indicate which extensions of the engine would increase the coverage of the collection (and likely of other published databases) the most (\Cref{fig:corpus}b). Missing phases most often require the compound energy formalism with more than one sublattice (\rerun{\num{912}} databases), which is used for ordered intermetallic phases such as B2, Laves, and sigma. Other missing phases are interstitial solutions with a sublattice of vacancies (\rerun{\num{351}} databases), phases with charged or other non-element species, e.g., ionic liquids (\rerun{\num{199}}), and gas phases (\rerun{\num{196}}), while \rerun{\num{49}} databases do not give the Gibbs energy of a pure end-member for at least one phase.  Physics is dropped mostly because of associate species (\rerun{\num{187}} phases), magnetic data that cannot be evaluated (\rerun{\num{121}}), terms that the parser cannot read (\rerun{\num{82}}), and parameter types of the third-generation unary descriptions (\rerun{\num{56}}).

\begin{figure*}[t]
  \centering
  \includegraphics[width=\textwidth]{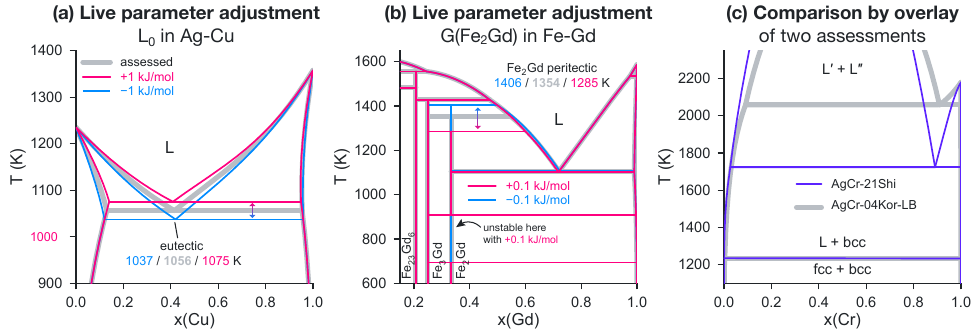}
  \caption[Living]{Sensitivity of phase diagrams to model parameters and differences between assessments, computed with \tool. (a) Ag--Cu \cite{hayes1986} with the zeroth-order parameter $L_0$ of the liquid changed by \href{https://gibbscope.github.io/?db=AgCu-86Hay&x=0,1&t=900,1400&g=0&fig=light&mode=vector&p=LIQUID:0:a=18534.6}{\qty{+1}{\kilo\joule\per\mole}} (pink) and \href{https://gibbscope.github.io/?db=AgCu-86Hay&x=0,1&t=900,1400&g=0&fig=light&mode=vector&p=LIQUID:0:a=16534.6}{\qty{-1}{\kilo\joule\per\mole}} (blue) from the assessed value (thick gray), which moves the eutectic from \rerun{\qty{1056}{\kelvin}} to \rerun{\num{1075}} and \rerun{\qty{1037}{\kelvin}}, respectively. (b) Fe--Gd \cite{chen2017} with the Gibbs energy of Fe$_2$Gd changed by \href{https://gibbscope.github.io/?db=FeGd-17Che&x=0.15,1&t=600,1650&g=0&fig=light&mode=vector&p=_compound:FE2GD_C15=300}{\qty{+0.1}{\kilo\joule\per\mole}} (pink) and \href{https://gibbscope.github.io/?db=FeGd-17Che&x=0.15,1&t=600,1650&g=0&fig=light&mode=vector&p=_compound:FE2GD_C15=-300}{\qty{-0.1}{\kilo\joule\per\mole}} (blue), which moves its peritectic formation from \rerun{\qty{1354}{\kelvin}} to \rerun{\num{1285}} and \rerun{\qty{1406}{\kelvin}}, respectively; with \qty{+0.1}{\kilo\joule\per\mole}, Fe$_2$Gd is unstable between \rerun{\num{693}} and \rerun{\qty{907}{\kelvin}}. (c) Overlay of two assessments of Ag--Cr in \tool (\href{https://gibbscope.github.io/?db=AgCr-21Shi&bl=AgCr-04Kor-LB&x=0,1&t=1100,2350&g=0&fig=light&mode=vector}{view online}), AgCr-21Shi \cite{shi2021phase} (purple) and AgCr-04Kor-LB \cite{korb2004gtt} (thick gray, loaded with ``Compare with''), whose monotectic temperatures differ by \rerun{\qty{335}{\kelvin}}. For this figure, phase boundaries were computed with the \tool engine and replotted for clarity; each hyperlinked value opens the same state in \tool in its format.}
  \label{fig:living}
\end{figure*}

Calculation of phase diagrams for such a large collection also tests the computational efficiency of \tool at scale. For the entire bundled collection, it took \tool (running in eight parallel processes) \rerun{\qty{182}{\second}} to compute phase diagrams with the default settings of the application: 200 temperature steps and a \qty{0.5}{\kelvin} bracket. As expected, the time needed for a phase diagram increases with the number of phases computed (\Cref{fig:corpus}c). The median time across the collection was \rerun{\qty{0.35}{\second}}, and the maximum was \rerun{\qty{1.22}{\second}} per phase diagram -- all obtained on a consumer-grade laptop (see \Cref{sec:methods-timing}). Across the collection, the engine found \rerun{\num{11317}} invariant reactions, including the most common eutectic (\rerun{\num{3939}}) and peritectic (\rerun{\num{3517}}) reactions. Only two changes in the sequence of stable phases in the entire collection could not be classified at the default settings; both are eventually resolved with a finer temperature bracket of \qty{0.1}{\kelvin}.

\subsection{Sensitivity of phase diagrams to model parameters and assessments}
\label{sec:living}

The speed of the engine allows us to examine the two aspects of living phase diagrams discussed in \Cref{sec:intro}: (i) the sensitivity of a phase diagram to the parameters of a given assessment and (ii) the differences between assessments of the same system. Both aspects can be explored interactively in the application with parameter sliders and the overlay of assessments (\textcircled{4}--\textcircled{5} in \Cref{fig:concept}; \Cref{fig:living}). Here, we demonstrate analyses enabled by these features on example systems.

To quantify the sensitivity of phase diagrams to selected model parameters, we changed the constant term of each Redlich--Kister parameter by \qty{+-1}{\kilo\joule\per\mole} and the Gibbs energy of each line compound by \qty{+-0.1}{\kilo\joule} per mole of atoms, one at a time, and computed the resulting changes in invariant temperatures by central differences. The analysis was performed for assessments of four familiar systems with one or two invariant reactions: Ag--Cu \cite{hayes1986} (\Cref{fig:living}a), Al--Si \cite{feufel1997investigation}, Al--Zn \cite{mey1986thermodynamic}, and Bi--Sn \cite{ohtani1994thermodynamic} (datasets AlSi-97Feu, AlZn-93Mey, and BiSn-94Oht of the SGTE collection) and for the assessment of the Fe--Gd system \cite{chen2017} (\Cref{fig:living}b). In the four simple systems, eutectic temperatures are most sensitive to the zeroth-order (regular-solution) parameter of the liquid: a change of \qty{1}{\kilo\joule\per\mole} in this parameter shifts the eutectic temperature by \rerun{\qty{19}{\kelvin}} in Ag--Cu, \rerun{\qty{13}{\kelvin}} in Bi--Sn, \rerun{\qty{9}{\kelvin}} in Al--Zn, and \rerun{\qty{7}{\kelvin}} in Al--Si. The largest sensitivity was found for the monotectoid reaction in Al--Zn, which terminates the miscibility gap in the f.c.c.\ phase and shifts by \rerun{\qty{46}{\kelvin}} per kJ/mol of the zeroth-order parameter of the f.c.c.\ phase. Invariant reactions involving line compounds in Fe--Gd are even more sensitive. The peritectic formation of Fe$_2$Gd (\rerun{\qty{1354.5}{\kelvin}}) shifts by \rerun{\qty{60}{\kelvin}} per \qty{0.1}{\kilo\joule\per\mole} (per mole of atoms) change in the Gibbs energy of Fe$_2$Gd. An increase of \qty{0.1}{\kilo\joule\per\mole} in the Gibbs energy of Fe$_2$Gd also changes the sequence of reactions: Fe$_2$Gd decomposes into Fe$_3$Gd and the h.c.p.\ phase in a eutectoid reaction at \rerun{\qty{906.5}{\kelvin}} and forms again in a peritectoid reaction at \rerun{\qty{692.8}{\kelvin}}, so that it is unstable between these temperatures (\Cref{fig:living}b). These changes in model parameters are small compared with the typical uncertainty of measured enthalpies of formation of intermetallic compounds, which is of the order of \qty{1}{\kilo\joule} per mole of atoms \cite{kim2017experimental} (calorimetric measurements of the same compound in different studies scatter by about \qty{3}{\kilo\joule} per mole of atoms \cite{kim2017experimental,arroyave2026orchid}). The step of \qty{0.1}{\kilo\joule} per mole of atoms for compounds is an order of magnitude smaller, and a change of \qty{1}{\kilo\joule\per\mole} in the constant term of a Redlich--Kister parameter changes the enthalpy of mixing by at most \qty{0.25}{\kilo\joule\per\mole}.

The comparison of assessments facilitated by \tool is illustrated (\Cref{fig:living}c) on two databases of Ag--Cr \cite{shi2021phase,korb2004gtt} from the SGTE collection (AgCr-21Shi and AgCr-04Kor-LB). In both assessments, the liquid has a miscibility gap, and the Cr-rich liquid decomposes on cooling into the Ag-rich liquid and b.c.c.\ Cr in a monotectic reaction. This reaction, however, is at \rerun{\qty{1723}{\kelvin}} in \cite{shi2021phase} and at \rerun{\qty{2058}{\kelvin}} in \cite{korb2004gtt}, i.e., \rerun{\qty{335}{\kelvin}} apart, whereas the two diagrams nearly coincide at lower temperatures, including the reaction near the melting point of Ag. Such a difference is more than an order of magnitude larger than the scatter of invariant temperatures measured in different studies (about \qty{10}{\kelvin} \cite{arroyave2026orchid}) and becomes apparent as soon as the second assessment is loaded using the ``Compare with'' feature of the \tool interface.

These examples show that invariant temperatures respond strongly to changes in model parameters that are well within the uncertainty of the underlying thermochemical data and that different assessments of the same system can place an invariant reaction more than \qty{300}{\kelvin} apart. A static figure of a single assessment conveys neither of these aspects, while \tool offers a convenient and interactive route for their exploration. 

\section{Discussion}
\label{sec:discussion}

\subsection{CALPHAD interactivity and its implications for education and research}

The results presented above show that complete CALPHAD calculations of binary phase diagrams can be performed in a web browser at interactive speed without compromising accuracy. The \tool engine reproduces PyCalphad and Thermo-Calc to within millikelvins in temperature and \num{e-6} in composition (\Cref{sec:validation}), handles every database of the SGTE collection with an explicit report of unsupported phases (\Cref{sec:corpus}), and recomputes a phase diagram (upon parameter change) in a fraction of a second (\Cref{sec:framework}). This combination of accuracy, coverage, and speed allows phase diagrams to be used in new ways that are impractical with static figures.  We envision at least four contexts that can benefit from this interactivity.

First, for teaching, \tool displays the Gibbs energy curves of all phases and their lower convex hull together with the phase diagram on the same composition axis (\textcircled{3} in \Cref{fig:concept}). Students can follow how a common tangent becomes a two-phase field as the temperature changes and how a change in an interaction parameter of the liquid shifts the eutectic point. While textbook figures can  show these evolutions only at a few selected temperatures, GibbScope displays these changes continuously, which can enhance learning similar to interactive simulations \cite{wieman2008phet}. Since the application requires neither installation nor a paid license, it can be used directly in classes and assignments. 

Second, for reviewing and selecting assessments, the overlay of assessments and the per-point residuals allow users to compare competing descriptions of the same system against the same experimental data (\Cref{fig:concept,fig:living}c). As the example of Ag--Cr shows (\Cref{sec:living}), assessments of the same system can differ by far more than the uncertainty of experiments, which makes such a comparison a useful first step in selecting a thermodynamic database. Experimental phase equilibria for the systems of the SGTE collection have recently been extracted from the literature and released as ESPEI datasets \cite{arroyave2026orchid}, which could make such comparisons routine.

Third, for developing thermodynamic databases, \tool can serve as a preliminary step before a full optimization with dedicated tools (e.g., PARROT module of Thermo-Calc \cite{andersson2002} or ESPEI \cite{bocklund2019}). The effect of individual parameters can be explored interactively, problems in the experimental data can be identified from the residuals, and the modified database can be exported in the \texttt{.tdb} format for further work. Watching how a phase diagram responds to changes in individual parameters is valuable for beginner CALPHAD practitioners to develop a ``feel'' for how different features of the phase diagram and its topology depend on the individual parameters.  This understanding  is  difficult to develop otherwise, with advanced optimization software that adjusts multiple parameters at once in more of a ``black-box'' fashion. 

Fourth, for communicating results, links generated by \tool reopen the same views of a database from the collection, including modified parameters (as in \Cref{fig:living}). A phase diagram in a paper could therefore be accompanied by a link that allows the reader to examine the underlying model and zoom into different regions of interest in a phase diagram instead of a single static snapshot, much as interactive articles \cite{hohman2020communicating} and computational notebooks \cite{kluyver2016jupyter,ziatdinov2019building} give readers access to the models and data behind a figure. The current application could serve as a foundation for a centralized platform for publication, analysis, and sharing of assessments driven by the CALPHAD community, as envisioned for CALPHAD data repositories \cite{campbell2014file} and for FAIR materials data more broadly \cite{scheffler2022fair}.

\subsection{Limitations}

The present implementation of \tool has several limitations. First, the engine is limited to binary systems with substitutional solution phases and line compounds. Phases described with the compound energy formalism (including ordered phases with order--disorder models), ionic liquids, gas phases, and third-generation descriptions of the unaries are not supported. As a result, \rerun{\num{1343}} of the \rerun{\num{2745}} databases in the collection have at least one phase missing (\Cref{sec:corpus}) and the resulting phase diagrams are incomplete. Second, the resolution of the calculations is finite. A miscibility gap narrower than \num{0.004} in mole fraction is treated as a single phase, so that a gap narrowing slowly towards its critical point is cut below the true critical point. This is the case for \rerun{\num{130}} of the \rerun{\num{388}} critical points in the collection; for example, the true critical point of the f.c.c.\ gap in NiTi-15San lies about \rerun{\qty{55}{\kelvin}} higher. Phase fields narrower in temperature than one step of the scan can be missed (e.g., the \qty{0.2}{\kelvin} field of Co$_5$Ge$_2$ in Co--Ge), and solubilities below about \num{e-8} are read as zero. These numerical settings were chosen as a compromise between accuracy and speed and can be refined at a higher computational cost. Third, the fitting in \tool minimizes the distance between experimental points and the computed diagram scaled by the uncertainties of the points. Unlike the driving-force formulation of phase-equilibrium data used in ESPEI \cite{bocklund2019}, this objective does not account for the thermodynamic driving force, and the fitted parameters come without uncertainty estimates. The fitting in \tool therefore serves as a means of exploration and preliminary analysis, while full assessments should be performed with dedicated software using the exported database.

Several extensions of \tool can naturally follow from these limitations and from the results presented above. Implementation of the compound energy formalism would recover the phases currently missing in \rerun{\num{1343}} databases of the collection, most of which are ordered intermetallic and interstitial phases (\Cref{sec:corpus}). Extension to ternary systems, starting with isothermal sections, would bring the approach closer to the multicomponent alloys of practical interest. Since the engine computes a phase diagram in a fraction of a second, it could also visualize the results of Bayesian uncertainty quantification (e.g., \cite{paulson2019,honarmandi2019}) directly in the browser, i.e., display distributions of phase boundaries computed for an ensemble of parameter sets instead of a single diagram. 

\subsection{Robust development with agentic AI}

\tool was developed with the assistance of AI coding agents (Claude Opus 5.x), which handled bulk and routine code writing under the direction of the authors.
We therefore treated verification of the code as an integral part of the development. Many of the numerical results and statements about the engine rely on automated checks: the comparisons with PyCalphad and Thermo-Calc (\Cref{sec:validation}), the loading and analysis of the entire SGTE collection (\Cref{sec:corpus}), comparisons of boundary calculations for \rerun{\num{38}} systems with captured references, \rerun{\num{318}} checks of the compiled WebAssembly module, and \rerun{\num{699}} assertions on the rendered application (documented in the code repository). 

\section{Conclusion}

We present \tool, a freely available framework for interactive calculation and visualization of binary phase diagrams in the browser. Written in Rust and compiled to WebAssembly, the engine reproduces the results of established CALPHAD software to within numerical precision, handles the SGTE collection of binary datasets with an explicit account of unsupported phases, and is fast enough for the phase diagram to follow changes in model parameters as they are made. Using this speed, we showed that assessed phase diagrams are sensitive to changes in parameters well within the uncertainty of the underlying data and that assessments of the same system can differ by far more than the uncertainty of experiments. These results show the potential of browser-based CALPHAD calculations for making phase diagrams live objects that can be examined, compared, and revised  by not only thermodynamic database developers but also their readers. This approach could extend to other materials applications offering a shift from static snapshots to explorable models. 

\section*{Code availability}

The application is freely available at \url{https://gibbscope.github.io}. The source code of the engine and the application will be made available at \url{https://github.com/GibbScope/gibbscope}.

\section*{Data availability}

The present analysis and \tool use the SGTE collection of binary datasets originally published by Hallstedt \cite{hallstedt2025}, included in the application unmodified under the CC BY 4.0 license.  The experimental phase-boundary data of Cr--Ni shown in \Cref{fig:concept} and loaded with this example in the application were adopted from Tang and Hallstedt \cite{tang2016using}, which compiles measurements from the original studies cited therein. The experimental data bundled with other preloaded examples are credited to their original sources in the application.

\section*{Acknowledgments}

This work was supported by the National Science Foundation under Award No.\ 2441813. 

\section{Methods}
\label{sec:methods}

\subsection{Thermodynamic models}
\label{sec:methods-models}

The \tool engine computes phase equilibria in binary systems A--B at constant pressure. The composition is expressed by the mole fraction $x$ of element B, which is the second element declared in the database, so that $x_\mathrm{A} = 1 - x$ and $x_\mathrm{B} = x$. Solution phases are described as substitutional solutions of the two elements on one sublattice, and their Gibbs energy per mole of atoms is given by
\begin{equation}
\label{eq:gibbs}
  \begin{split}
  G^{\varphi} &= \frac{1}{a} \times \\
  & \times \left[x_\mathrm{A}\,{}^{\circ}G_\mathrm{A}^{\varphi} + x_\mathrm{B}\,{}^{\circ}G_\mathrm{B}^{\varphi} + x_\mathrm{A}x_\mathrm{B}\sum_{\nu} L_{\nu}^{\varphi}\,(x_i - x_j)^{\nu}\right]\\
  &+RT\left(x_\mathrm{A}\ln x_\mathrm{A} + x_\mathrm{B}\ln x_\mathrm{B}\right) + G_\mathrm{mag}^{\varphi},
  \end{split}
\end{equation}
where ${}^{\circ}G_\mathrm{A}^{\varphi}$ and ${}^{\circ}G_\mathrm{B}^{\varphi}$ are the Gibbs energies of the pure elements in the structure of phase $\varphi$ referred to the SGTE unary data \cite{dinsdale1991}, $L_{\nu}^{\varphi}$ is the Redlich--Kister parameter of order $\nu$ \cite{redlich1948}, $R = \qty[per-mode=power]{8.31451}{\joule\per\mole\per\kelvin}$, and $G_\mathrm{mag}^{\varphi}$ is the magnetic contribution. The factor $a$ is the number of sites of the mixing sublattice per formula unit (usually $a = 1$), which converts the end-member and excess terms given per formula unit in the database into energies per mole of atoms, while additional sublattices occupied only by vacancies (e.g., in $(\mathrm{A},\mathrm{B})_1(\mathrm{Va})_3$) contribute no atoms. In \Cref{eq:gibbs}, $i$ and $j$ denote the two elements in alphabetical order regardless of the order in which they are written in a parameter or declared in the database. This rule sets the sign of the odd-order terms in the same way as in PyCalphad and Thermo-Calc. The unary Gibbs energies and the Redlich--Kister parameters are piecewise functions of temperature, and each piece is a linear combination of $1$, $T$, $T\ln T$, $T^2$, $T^3$, $T^{-1}$, $T^7$, $T^{-9}$, and other integer powers of $T$ (if present in the database). Below the lowest temperature range of a function, its first piece is extended to lower temperatures, and above the highest range, its last piece is used.

The magnetic contribution follows the model of Inden \cite{inden1981} in the form given by Hillert and Jarl \cite{hillert1978}:
\begin{equation}
  G_\mathrm{mag}^{\varphi} = RT\ln(\beta + 1)\,g(\tau), \qquad \tau = \frac{T}{T_\mathrm{C}},
  \label{eq:magnetic}
\end{equation}
where $\beta$ is the mean magnetic moment in Bohr magnetons, $T_\mathrm{C}$ is the critical temperature of magnetic ordering (Curie or N\'eel temperature), and $g(\tau)$ is the polynomial of Hillert and Jarl with separate branches below and above $T_\mathrm{C}$. The structure factor $p$ in $g(\tau)$ (0.4 for b.c.c.\ and 0.28 for f.c.c.\ and h.c.p.\ phases) and the antiferromagnetic factor ($-1$ for b.c.c.\ and $-3$ for f.c.c.\ and h.c.p.\ phases) are read from the \texttt{TYPE\_DEFINITION} statement of the phase. The composition dependence of $T_\mathrm{C}$ and $\beta$ is described by Redlich--Kister expansions of the \texttt{TC} and \texttt{BMAGN} parameters, and negative values of these quantities are divided by the antiferromagnetic factor following the implementation in PyCalphad.

Line compounds are phases in which each sublattice is occupied by a single element or by vacancies. The composition of a line compound is calculated from the site ratios of its sublattices, counting only the sites occupied by atoms. Its Gibbs energy is that of its end-member divided by the number of atoms per formula unit and includes the magnetic contribution if the database gives one. In the application, the user can change the constant term (A) of each Redlich--Kister parameter that describes the interaction of the two elements, i.e., $L_{\nu}^{\varphi} = A + BT + \ldots$, and add a constant offset to the Gibbs energy of each line compound (analyzed in \Cref{fig:living}b for Fe$_2$Gd). Such a change replaces only the coefficient A (or adds the offset), while all other terms given in the database, including the temperature-dependent ones, are retained. 

\subsection{Database parsing and phase grading}
\label{sec:methods-parsing}

The parser reads a \texttt{.tdb} file in three stages. First, comments (from \texttt{\$} to the end of a line) are removed, and the text is split into statements terminated by an exclamation mark. Second, the type of each statement (e.g., \texttt{ELEMENT}, \texttt{FUNCTION}, \texttt{PHASE}, \texttt{CONSTITUENT}, \texttt{PARAMETER}, or \texttt{TYPE\_DEFINITION}) is identified, and its contents are stored. Third, references between functions are resolved by substituting the polynomial coefficients of each referenced function into the expression that refers to it. Products of functions and powers of $T$ are expanded over the intersection of their temperature ranges as long as the result remains a polynomial of the form given in \Cref{sec:methods-models}. Terms that cannot be written in this form (e.g., an exponential function of $T$) and references to functions that the database does not define are recorded as unreadable terms of the phase instead of being silently set to zero.

As a result, every Gibbs energy expression is stored in memory as a list of polynomial coefficients for each temperature range. Evaluation of the Gibbs energy at a given composition and temperature therefore requires only a small number of arithmetic operations, and a change of a parameter in the application replaces one coefficient without parsing the database again. This approach differs from that of PyCalphad, which constructs symbolic expressions of the Gibbs energy with a computer algebra system and compiles them before evaluation \cite{otis2024pycalphad}. The present approach is less general, but it is sufficient for the models described in \Cref{sec:methods-models} and avoids the computational cost of the symbolic step.

After parsing, each phase declared in the database is graded. A phase is computed as a solution phase if it has exactly one sublattice on which constituents mix, both elements are present on this sublattice, all other sublattices contain only vacancies, and the database gives the Gibbs energies of both pure end-members. A phase is computed as a line compound if each of its sublattices contains a single constituent, all its constituents are elements of the system, it is not a gas, and the database gives the Gibbs energy of its end-member. Any other phase is graded \textit{phase missing} and left out of the calculation. The reason reported for a missing phase is determined in the following order: (i) a gas phase; (ii) an interstitial solution, i.e., an element mixing with vacancies on a sublattice of its own; (iii) constituents other than the two elements (e.g., ions or molecules); (iv) more than one sublattice occupied by atoms, which requires the compound energy formalism; and (v) a missing Gibbs energy of a pure end-member. A computed phase is graded \textit{physics dropped} if the database gives a contribution that the engine does not evaluate, namely (i) associate species, which are dropped so that the phase is computed as a substitutional solution of the two elements; (ii) unreadable terms identified during parsing; (iii) parameter types that are not implemented, which are currently those of the third-generation unary descriptions (\texttt{THETA}, \texttt{LNTHETA}$n$, \texttt{THETAF}$n$, and \texttt{GD}) and any other type that the parser does not recognize; or (iv) magnetic data that cannot be evaluated, i.e., a critical temperature without a magnetic type definition, the N\'eel temperature of the third-generation magnetic model \cite{xiong2012improved}, or a magnetic parameter given twice. All other computed phases are graded \textit{modeled in full}. Phases excluded by \texttt{REJECT\_PHASE} statements are graded \textit{rejected by the database}, and a database in which no phase can be computed is refused with an explicit reason. Each grade is stored together with a machine-readable reason and a description of the omitted contribution, which the application displays in the fidelity panel (\textcircled{8} in \Cref{fig:concept}) and in the caption of the phase diagram. A database takes the grade of its least supported phase, i.e., \textit{modeled in full}, \textit{physics dropped}, or \textit{phases missing}.

\subsection{Calculation of phase equilibria at a given temperature}
\label{sec:methods-hull}

At a given temperature and pressure, the equilibrium state of a binary system minimizes its Gibbs energy, which corresponds to the lower convex hull of the Gibbs energy curves of all phases \cite{gibbs1876}. To construct this hull, each solution phase is sampled on a uniform grid of \num{1001} compositions ($\Delta x = \num{e-3}$) and, in addition, at 20 compositions near each pure element, spaced logarithmically (four per decade) at distances from \num{e-8} to \num{5.6e-4} from the pure element. The additional points are needed because a terminal solid solution can be stable entirely within the first interval of the uniform grid. Each line compound contributes a single point at its composition. The lower convex hull of all points is computed with the monotone chain algorithm \cite{andrew1979another}, and each vertex of the hull retains the phase that it belongs to. A segment of the hull that connects two different phases is a tie line of a two-phase equilibrium. A segment that connects two points of the same phase more than \num{4e-3} apart (i.e., four steps of the uniform grid) is a tie line across a miscibility gap in that phase, while consecutive vertices of the same phase closer than this threshold belong to a single-phase region.

Since the vertices of the hull are sampled points, the compositions of coexisting phases obtained from the uniform grid are accurate only to the grid spacing. The hull is therefore refined in four passes. In each pass, the phases are sampled with eight times finer spacing 
on either side of (i) each end of a tie line, (ii) the ends of same-phase segments that may be narrow miscibility gaps, and (iii) each location where a phase dips below the hull between its own samples, after which the hull is computed again. After four passes, the local spacing is $10^{-3}/8^4 \approx 2.4\times10^{-7}$, which is finer than the precision of the reference data used for validation (\Cref{sec:validation}). Different fields of the same phase (e.g., the Ag-rich and Cu-rich f.c.c.\ solutions in Ag--Cu) are distinguished as composition sets numbered in the order of their mean composition. 

\subsection{Temperature scan and invariant reactions}
\label{sec:methods-sweep}

The temperature range of a phase diagram is determined from the database before the scan: the lower limit is the lowest temperature declared in the database, and the upper limit is placed above the last solidus. The phase diagram is then computed in a scan over 200 equal temperature steps. At each of the resulting 201 temperatures, the hull gives the sequence of stable phase fields from $x = 0$ to $x = 1$, i.e., the single-phase regions and tie lines in the order of composition. Where the sequences at two neighboring temperatures differ, the temperature interval is bisected until its width is below \qty{0.5}{\kelvin}, and the change in the sequence is classified. 

A change in which one tie line is replaced by two tie lines that meet at a third phase, or vice versa, is a three-phase invariant reaction. Once a change is classified as an invariant reaction, its bracket is bisected six more times (i.e., to 1/64 of its width). The temperature of the reaction is taken as the midpoint of the final bracket, and the composition of each participating phase is taken as the mean of its values at the two ends of the bracket, with half of their difference reported as the uncertainty. The tie-line ends at all temperatures of the scan and bisection are connected into boundary curves of two-phase fields. A miscibility gap whose width falls below the threshold of \num{4e-3} while the gap is still open \qty{1}{\kelvin} above the last resolved temperature is marked as ``narrowed below'' its critical point.

Invariant reactions are named after the middle phase, i.e., the phase that is consumed or formed on cooling, and the two outer phases. If the middle phase decomposes into the outer phases on cooling, the reaction is eutectic ($\mathrm{L} \rightarrow \alpha + \beta$), monotectic ($\mathrm{L}_1 \rightarrow \alpha + \mathrm{L}_2$), metatectic ($\beta \rightarrow \alpha + \mathrm{L}$), eutectoid ($\gamma \rightarrow \alpha + \beta$), or monotectoid ($\alpha_1 \rightarrow \alpha_2 + \beta$), depending on which phases are liquid and whether one phase participates at two compositions. If the middle phase forms from the outer phases on cooling, the reaction is peritectic ($\mathrm{L} + \alpha \rightarrow \beta$), syntectic ($\mathrm{L}_1 + \mathrm{L}_2 \rightarrow \alpha$), or peritectoid ($\alpha + \beta \rightarrow \gamma$). Here, L denotes a phase named as a liquid in the database. The engine also identifies degenerate reactions, in which one phase replaces another at a single composition. If this composition is that of a compound with two forms, the reaction is classified as polymorphic. If the reaction occurs at a pure element, it is named according to which of the two forms of the element dissolves more of the other element, which is judged by the quantity $k = \mu - RT\ln x$ of the dilute element in each form at the reaction temperature; the reaction is classified as allotropic if the two values of $k$ agree within \qty{e-3}{\joule\per\mole}. Reactions that fit none of these types are labeled ``other''. Finally, the upper limit of the temperature ranges declared in the database is recorded for each phase. Temperatures above the lowest such limit among the phases on the hull are hatched in the phase diagram, and an invariant reaction is flagged as extrapolated if at least one of its participating phases is above its declared limit.

In addition to the boundary curves and invariant reactions, the scan returns a raster of $201\times201$ cells over the displayed range of compositions and temperatures. The raster is used to fill the phase fields, whereas phase boundaries and invariant reactions are drawn as vector lines. The number of temperature steps and the width of the bisection bracket are arguments of the engine, which are fixed at 200 and \qty{0.5}{\kelvin} in the application.

\subsection{Misfit to experimental data and fitting of parameters}
\label{sec:methods-misfit}

Each experimental point is described by its composition $x_k$, temperature $T_k$, type, uncertainties $\sigma_{x,k}$ and $\sigma_{T,k}$ (\num{0.01} and \qty{10}{\kelvin} by default), weight $w_k$ (1 by default), and optionally the phase it belongs to as well as its source. The types of points are \texttt{boundary} (including liquidus, solidus, and solvus points), \texttt{tie\_left}, \texttt{tie\_right}, \texttt{invariant}, and \texttt{single\_phase}. Points are imported from CSV, TSV, and JSON files or converted from PARROT \texttt{.pop} files and ESPEI datasets of phase equilibria. From \texttt{.pop} files, the tables that fix one composition and one temperature per row are read, with the uncertainty given in the experiment statement and the weight from the \texttt{SET-WEIGHT} statement, while thermochemical data (e.g., enthalpies and activities) are reported as not applicable. In ESPEI datasets, one measured composition gives a boundary point and two give the ends of a tie line, and the default uncertainties are applied since these datasets do not carry uncertainties. Phase names in both formats are matched to the phases of the loaded database either exactly or, following the convention of Thermo-Calc, by a unique prefix (e.g., \texttt{LIQ} for \texttt{LIQUID}).

The residual $r_k$ of a point is its distance to the nearest matching feature of the computed phase diagram in coordinates scaled by the uncertainties of the point:
\begin{equation}
  r_k = \min_{(x,T) \in \mathcal{F}_k} \sqrt{\left(\frac{x - x_k}{\sigma_{x,k}}\right)^2 + \left(\frac{T - T_k}{\sigma_{T,k}}\right)^2},
  \label{eq:residual}
\end{equation}
where $\mathcal{F}_k$ is the set of features against which point $k$ is scored. For boundary and tie-line points, $\mathcal{F}_k$ consists of all segments of the boundary curves or, if the point names a phase, of the boundary curves of that phase. For invariant points, $\mathcal{F}_k$ consists of the compositions of the phases participating in invariant reactions at the temperatures of these reactions, so that a thermal arrest is compared with the nearest participating phase and not with the horizontal line of the reaction. A single-phase point has a zero residual if the computed diagram has the named phase (or any single phase if no phase is named) at $(x_k, T_k)$, and otherwise it is scored against the nearest boundary curve. If the diagram has no feature against which a point can be scored, the point is assigned a fixed residual of 50, so that removing a phase from the diagram cannot decrease the misfit. Points outside the temperature range declared in the database are reported but not scored. The misfit is the weighted root mean square of the residuals,
\begin{equation}
  S = \left(\frac{\sum_k w_k r_k^2}{\sum_k w_k}\right)^{1/2},
  \label{eq:misfit}
\end{equation}
which is also reported separately for each type of point. 

Selected parameters are fitted to the experimental data by minimizing $S$ with the Nelder--Mead simplex method \cite{nelder1965simplex}. A derivative-free method was chosen because $S$ is not differentiable wherever a change of parameters changes the topology of the phase diagram. The parameters unlocked by the user (constant terms of Redlich--Kister parameters and offsets of the Gibbs energies of line compounds) are scaled as $p = p_0 + u\,s$, where $p_0$ is the starting value, $u$ is the normalized variable, and $s = \max(0.05|p_0|, \qty{250}{\joule\per\mole})$. The search is constrained to the ranges of the corresponding sliders, i.e., \num{+-25}\% of the assessed coefficient but at least \qty{+-5000}{\joule\per\mole} for Redlich--Kister parameters and \qty{+-10000}{\joule} per formula unit for line compounds. Each trial vertex is moved into these bounds before evaluation instead of penalizing the objective, and parameters that end at a bound are reported to the user. The search stops when the relative difference of $S$ between the best and the worst vertex falls below \num{e-4}, when the size of the simplex falls below \num{e-4} in normalized units, or after 160 evaluations of $S$. Since each evaluation requires a complete temperature scan, the scans during fitting use 60 temperature steps instead of 200, which is still sufficient to separate invariant reactions about \qty{25}{\kelvin} apart, while invariant temperatures are bracketed to \qty{0.5}{\kelvin} as in the displayed diagram. The fitted parameters are applied only after the user accepts them.

\subsection{Implementation of the engine and the web application}
\label{sec:methods-implementation}

The source code of the engine is written in Rust \cite{matsakis2014rust}. The engine depends only on wasm-bindgen \cite{wasmbindgen} and the serde libraries for data serialization. It is compiled with wasm-pack 0.13.1 in release mode (optimization level 3 with link-time optimization), which also generates a JavaScript module that loads the WebAssembly binary and converts function arguments and results between the two languages. Data cross the interface as JSON strings (e.g., parameters, experimental points, and reports) and typed arrays (e.g., the raster of phase fields and Gibbs energy curves). After a database is loaded, the engine returns a handle object to JavaScript that holds the parsed database together with the evaluation plan of each phase (i.e., the coefficients of all its terms), so that subsequent calculations do not repeat the parsing.

The web application is written in HTML, CSS, and JavaScript. At page load, the main thread fetches and initializes the WebAssembly module, after which each user action (e.g., a change of a parameter or the temperature) calls the engine and redraws the figure(s). The phase diagram and the Gibbs energy curves are drawn on HTML canvas elements at the resolution of the display. For export in SVG, the same drawing code is run on a recording object that accepts the canvas drawing commands used by the application and writes each of them as an SVG element instead of painting pixels. The exported and the displayed figures are drawn by the same code  and thus show the same content. The temperature scan and the fitting run in two Web Workers, each holding its own instance of the engine and its own copy of the parsed database, since a handle created in one instance of the engine cannot be passed to another. 

The SGTE collection \cite{hallstedt2025} is served unmodified as static files together with a manifest of the phase grades of every database generated by the engine. The periodic table reads the manifest to display the grades without loading the databases. Only the database selected by the user is downloaded. The state of the view is encoded in the query string of the URL, which stores the values that differ from the defaults, i.e., the database and the compared assessment, the ranges of the axes, the temperature of the isotherm, the units and display settings, and the changed parameters (e.g., \texttt{p=LIQUID:0:a=19534.6}). Experimental data and the set of unlocked parameters are stored instead in a session file. For export of a modified database, the original file is split into statements, and only the \texttt{PARAMETER} statements of changed parameters are rewritten: a Redlich--Kister parameter as $A + BT + CT\ln T$ with twelve significant digits, and the Gibbs energy of a line compound by adding the offset to each of its temperature ranges. All other text is copied with the addition to a comment header that lists the changes and a reference entry for the changed parameters. The application is deployed as a static website on GitHub Pages.

\subsection{Verification}
\label{sec:methods-verification}

The verification of \tool combines tests of the engine against independent reference calculations, checks of the entire SGTE collection, and tests of the compiled WebAssembly module and the application (see the code repository for the latter). The Rust test suite includes unit tests of each module of the engine and comparisons with the following reference data. Invariant reactions of the curated set of 10 databases were computed with the mapping algorithm of PyCalphad 0.11.2 (\texttt{BinaryStrategy}). The phases in the PyCalphad calculations were restricted to those computed by the engine, so that the comparison measures numerical differences of supported phases instead of focusing on the extent of supported phase types. The tests require agreement within \qty{0.5}{\kelvin} in temperature and \num{5e-3} in composition (\qty{0.05}{\kelvin} for invariants that do not depend on the magnetic term), while the actual differences are reported in \Cref{sec:validation}. The magnetic contribution was compared point by point with that computed by PyCalphad. In addition, databases exported after a change of an interaction parameter were read with PyCalphad, and the resulting change of the Gibbs energy was compared with that applied by the engine. 

The comparison with PyCalphad was carried out on a random sample of 80 databases that the engine models in full. Each invariant reaction mapped by PyCalphad was matched to the nearest invariant reaction found by the engine with the same set of phases within \qty{3}{\kelvin}. The engine was run as the compiled WebAssembly module in Node.js at the settings of the application and again with 400 temperature steps and a bracket of \qty{0.1}{\kelvin} to distinguish reactions missed because of the resolution of the scan from disagreements between the two codes.

The comparison with Thermo-Calc was performed using console-mode macros generated based on the results of the engine. The macros skipped phases unsupported by \tool. The calculations were performed at $P = \qty{101325}{\pascal}$ for one mole of atoms. Each macro (i) mapped the phase diagram, (ii) solved each invariant reaction found by the engine by fixing one of its participating phases at zero amount and releasing the temperature, and (iii) computed equilibria at prescribed temperatures and compositions. In the tests, the engine is run with 400 temperature steps and a bracket of \qty{0.1}{\kelvin} and must reproduce every invariant reaction solved by Thermo-Calc with the same phases within \qty{0.01}{\kelvin} and \num{5e-5} in composition, report no additional invariant reactions, and find the same phases and tie-line ends within \num{e-6} at every prescribed temperature and composition.

\subsection{Measurement of calculation times}
\label{sec:methods-timing}

Calculation times were measured on an Apple M2 Pro computer (eight performance cores) running macOS, with the engine built from the release tag v1.0.0. In the browser, the time of one temperature scan (200 steps, a bracket of \qty{0.5}{\kelvin}, the full composition range, and the default temperature range) with the assessed parameters was measured in headless Chromium \rerun{149} driven by Playwright, and the median of \rerun{five} repetitions is reported for each database. The scan over the entire SGTE collection was performed with the same settings in Node.js \rerun{v24.13.0} with eight parallel processes, and the time of each database was recorded. These times include the calculation of the phase diagram but not the parsing of the database, which is performed once when a database is loaded.

\bibliography{refs}

\end{document}